# Magnetotransport in La(Fe,Ru)AsO as a probe of band structure and mobility

I.Pallecchi [1], F.Bernardini [2], M.Tropeano [1,5], A.Palenzona [1,3], A.Martinelli [1], C.Ferdeghini [1], M.Vignolo [1], S.Massidda [2], M.Putti [1,4]

[1] *CNR-SPIN, Corso Perrone 24, 16152 Genova, Italy*
[2] *Dipartimento di Fisica, Università di Cagliari, 09042 Monserrato (CA), Italy*
[3] *Dipartimento di Chimica e Chimica Industriale, Università di Genova, Via Dodecaneso 31, 16146 Genova, Italy*
[4] *Dipartimento di Fisica, Università di Genova, Via Dodecaneso 33, 16146 Genova, Italy*
[5] *Columbus Superconductors S.p.A, Via delle Terre Rosse 30, 16133 Genova, Italy*

**Abstract**
In this work we investigate the Ru substituted LaFeAsO compound, by studying the magnetotransport behaviour and its relationship with the band structure, in different regimes of temperature, magnetic field and Ru content. In particular we analyse the magnetoresistance of LaFe$_{1-x}$Ru$_x$AsO ($0 \leq x \leq 0.6$) samples with the support of *ab initio* calculations and we find out that in the whole series: (i) the transport is dominated by electron bands only; (ii) the magnetoresistance exhibits distinctive features related to the presence of Dirac cones; indeed, *ab initio* calculations confirm the presence of anisotropic Dirac cones in the band structure; (iii) the low temperature mobility is exceptionally high and reaches 18.6 m$^2$/(Vs) in the Ru-free sample at T=2K, in the extreme limit of a single Landau level occupied in the Dirac cones; (iv) the mobility drops abruptly above 10K-15K; (v) the disorder has a very weak effect on the band mobilities and on the transport properties; (vi) there exists a correlation between the temperature ranges of Dirac cones and SDW carrier condensation. These findings evidence the outstanding transport properties of Dirac cones in Fe-based pnictides parent compounds.

## 1. Introduction

With the discovery of superconductivity in 2008 by Kamihara [1] in LaFeAs(O,F), the REFeAs(O,F) family (RE=rare earth), also called 1111 family, has been devoted particular attention, as it exhibits the highest critical temperature among iron-based superconductors, Tc≈56K for RE=Sm [2]. The undoped parent compound is non superconducting and shows an antiferromagnetic/structural transition around 150K. Below this temperature, the ground state is a spin-density wave (SDW), arising from the nesting condition of different sheets of the Fermi surface. Doping destroys the SDW state and allows the superconducting state to arise. Since the earliest stages, it has been suggested that the nesting condition in the parent compound is a key element in triggering the superconducting pairing mechanism [3]. Indeed, it is believed that superconductivity is mediated by the antiferromagnetic spin fluctuations that form when the SDW state is suppressed. Hence, a close inspection of the band structure in the parent compound appears to be crucial in gaining full understanding of unconventional superconductivity in iron-based compounds. This task must be tackled from the theoretical point of view by *ab initio* and modelling approaches. In parallel, the expectations must be compared with experimental results. Beside spectroscopic measurements such as angle-resolved photoelectron spectroscopy (ARPES) [4], scanning tunnelling spectroscopy (STS) [5] and Shubnikov de Haas oscillations [6], the much simpler transport measurements may as well reveal the significant features of the band structure.

In order to better understand the interrelationship between magnetism, superconductivity and structural disorder in these compounds, we focus on the effect of Ru substitution in the parent compound LaFeAsO. It has been predicted that the isoelectronic Ru substitution at the Fe site in REFeAsO (RE=La and Sm) does not dope the system and the electronic structure is not significantly modified; in addition magnetic moment at the transition metal site is frustrated by the presence of the non magnetic substitution [7]. The suppression of the SDW by Ru has been indeed observed in PrFe$_{1-x}$Ru$_x$AsO compounds [8]. Moreover the investigation of the superconducting properties in SmFe$_{1-x}$Ru$_x$AsO$_{0.85}$F$_{0.15}$ has shown that for low level of substitution (0≤x<0.5) the effect of disorder induced by Ru dominates, while for large level of substitution (0.5≤x≤1) the changes in the band structure and the frustrated magnetism become significant [7].

Thus we want to tune the SDW transition temperature and the amount of disorder in La(FeRu)AsO compounds and use transport properties as a sensitive probe of the band structure. Indeed, some recent works [9,10,11] have shown that in the Ba(FeAs)$_2$ and PrFeAsO iron-based parent compound, two well discernible contributions to the magnetoresistance exist, namely a classic cyclotron one that allows to gain information on carrier density and mobility plus a positive linear contribution. The latter has been explained in terms of Dirac cones in the Fermi surface, following the model of quantum magnetoresistance developed by Abrikosov [12]. Dirac states are portions of the Fermi surface where the energy spectrum is linear in momentum. They are known to exist in the bulk [13] and surface [14] of topological insulators, in graphene [15], bismuth [16] and cuprate superconductors [17] and they are subject of intensive studies for their unique technological implications. Indeed, the effective mass associated to Dirac cone carriers is virtually zero, allowing high mobility transport and little effect of impurities.

In this work, we carry out measurements of resistivity, magnetoresistance and Hall effect on La(Fe,Ru)AsO samples and we analyze quantitatively our results with the support of *ab initio* calculations of the band structure. We evidence the presence of Dirac cones in the calculated band structure as well as in magnetotransport data and we explore their evolution by tuning the band structure, the SDW transition and the amount of disorder via Ru substitution.

**2. Experimental**
Polycrystalline LaFe$_{1-x}$Ru$_x$AsO (x=0, 0.1, 0.2, 0.3, 0.4, 0.5, 0.6) samples are prepared as described in ref.[7]. Structural characterization is performed by X-ray powder diffraction at room temperature and Rietveld refinement is carried out on selected diffraction patterns. Microstructure is observed by a polarized light microscopy (OM; Reichert-Jung, MeF3). For this aim, samples are embedded in cold-setting epoxy resin, cut and polished using up to 1 μm diamond paste. A final polishing stage is performed with an aqueous silica colloidal suspension. Transport properties measurements are carried out in a Physical Properties Measurement System (PPMS, Quantum Design) in the temperature range 2K-300K and in magnetic field up to 9T. Hall coefficients (R$_H$) are determined measuring the transverse resistivity at selected temperatures sweeping the field from -9T to 9T.

**3. Results**
Rietveld refinements of X-rays data reveal that all the samples crystallize in the tetragonal *P*4/*nmm* space group at room temperature; small amounts of La(OH)$_3$ (~7 vol%) can be detected. The refinement of the occupancies at the transition metal site reveals that the effective [Ru]/[Fe] ratio is in good agreement with the nominal one. Figure 1 shows the Rietveld refinement plot of the *x*=0.3 sample, selected as representative. As shown in the inset of figure 1, the cell edges exhibit a monotonic dependence on substitution and, similarly to what observed in the Sm(Fe$_{1-x}$Ru$_x$)As(O$_{0.85}$F$_{0.15}$) system [7], the *a* axis increases with Ru content whereas the *c* axis decreases.
Figure 2 shows a polarized OM image collected on the same *x*=0.3 sample. Different colour tones indicate different crystalline domains. Because of the high optical anisotropy characterizing these samples, the average domain size can be easily estimated, resulting on average 0.1-0.2 μm. The same average size is observed also in the other samples.

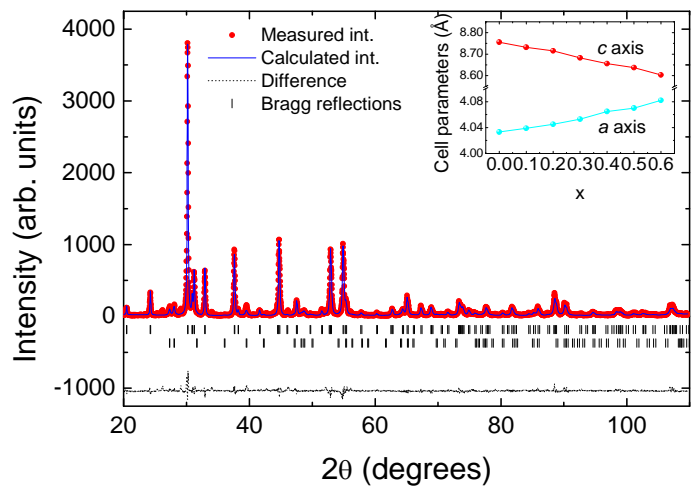

**Figure 1:** (Colour online) Rietveld refinement plot of La(Fe$_{0.7}$Ru$_{0.3}$)AsO; the inset shows the evolution of the unit cell parameters as a function of the Ru content.

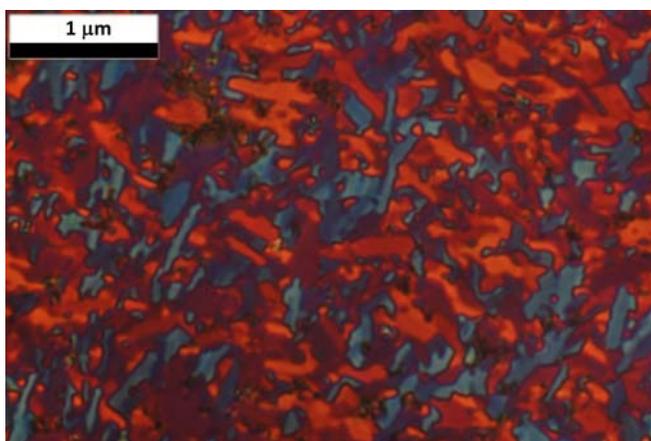

**Figure 2:** (Colour online) Microstructure (OM, polarized light) of La(Fe$_{0.7}$Ru$_{0.3}$)AsO, showing an average domain size in around 0.1-0.2 μm.

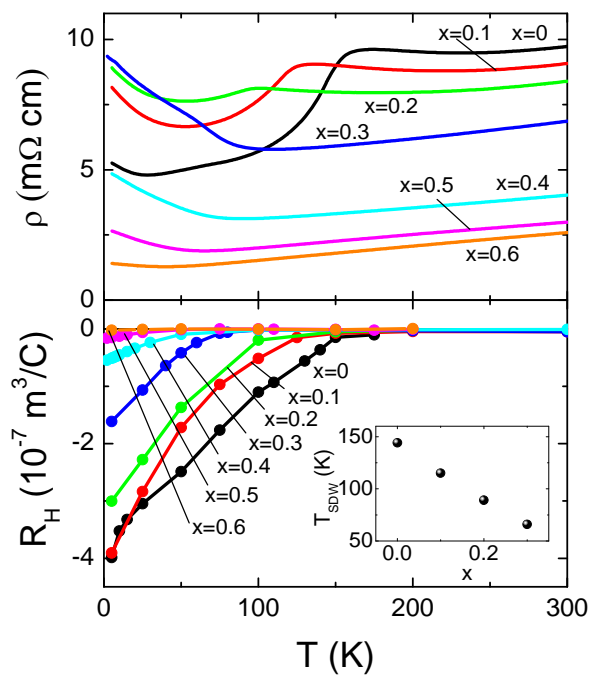

**Figure 3:** (Colour online) Upper panel: resistivity curves of the LaFe$_{1-x}$Ru$_x$AsO samples. Lower panel: Hall resistance curves of the same samples. Inset: onset temperature of the SDW transition as determined by the maximum of the resistivity derivative dρ/dT.

In the upper panel of figure 3, we present resistivity versus temperature curves of the whole series of LaFe$_{1-x}$Ru$_x$AsO samples. The x=0 sample exhibits a weak temperature dependence between 150K and room temperature, with ρ(300K)≈10 mΩ cm. At 150K, where the magnetic transition occurs, the resistivity drops abruptly and below 25K a resistivity upturn suggests carrier localization. With increasing Ru substitution, the room temperature resistivity decreases steadily down to a value ρ(300K)≈2.6 mΩ cm for the x=0.6 sample and consistently the high temperature metallic character is more evident with increasing x. These observations are in agreement with the phenomenology observed in Sm(Fe,Ru)As(O,F) samples [7], which has been explained in terms of *d* band broadening by Ru substitution. At the same time, the low temperature upturn becomes more evident with increasing Ru content for x from 0 to 0.3, while this localization effect decreases and almost disappears at higher substitution values. The magnetic transition is shifted to lower temperature by Ru substitution and disappears from the resistivity curves for x≥0.3.

In the lower panel of figure 3, the Hall resistance curves are plotted as a function of temperature for the whole series of samples. It can be seen that for all the samples the high temperature (T>200K) R$_H$ is very small and weakly temperature dependent, as expected for multiband almost compensated metals. With decreasing temperature, R$_H$ increases in magnitude and becomes more and more negative, indicating a predominance of electron-type carriers. The onset temperature for this departure coincides with the magnetic transition temperature in resistivity curves for the low x samples and decreases monotonically with increasing x up to x=0.6, where it is still well visible. This suggests that for x≥0.3 the magnetic transition is still present, even if resistivity curves are featureless. In the inset, the SDW transition temperature T$_{SDW}$, defined at the maximum of the resistivity derivative dρ/dT, is plotted as a function of the Ru content. Even if we have data points only up to x=0.3, the ever present onset in Hall resistance curves justifies a linear extrapolation of T$_{SDW}$ up to x=0.6. Furthermore, μSR measurements carried out on the same series of samples to probe the local magnetic ordering in the bulk confirm the linear trend of the SDW transition temperature as a function of x, with a still non vanishing temperature for x=0.5 [18].

In figure 4 we present magnetoresistance curves measured at different temperatures. For simplicity, we show only three samples, namely x=0 in the upper panel, x=0.2 in the second panel and x=0.5 in the third panel. In the respective insets more curves measured at the lowest temperatures (T=2, 3, 4, ...,10K) and lowest field (-1T<μ$_0$H<1T) are shown. It can be seen that at low temperature (T≤10K) the x=0 sample presents a sharp dip at low field and a linear behaviour at higher field (μ$_0$H>2-3T). At higher temperature (T≥15K) the dip disappears and the magnetoresistance assumes the typical H$^2$ dependence; at the same time, the linear behaviour shift to higher fields and is not seen anymore in our experimental magnetic field window [-9T,+9T]. The x=0.5 sample in the second panel shows a similar behaviour, but the dip at low temperature and low field becomes weaker and the magnetoresistance is reduced. The x=0.2 sample in the second panel is midway in terms of magnetoresistance values, but it differs from the x=0 and x=0.5 samples in that it presents no sharp dip at low temperature and low field. However, the high field linear behaviour is similar in all the samples. In the bottom panel of figure 4, the magnetoresistances of all samples measured at the lowest temperature T=2K are shown in logarithmic scale. It can be seen that with increasing x the magnitude of magnetoresistance decreases and the crossover between the different H regimes shifts to slightly higher fields.

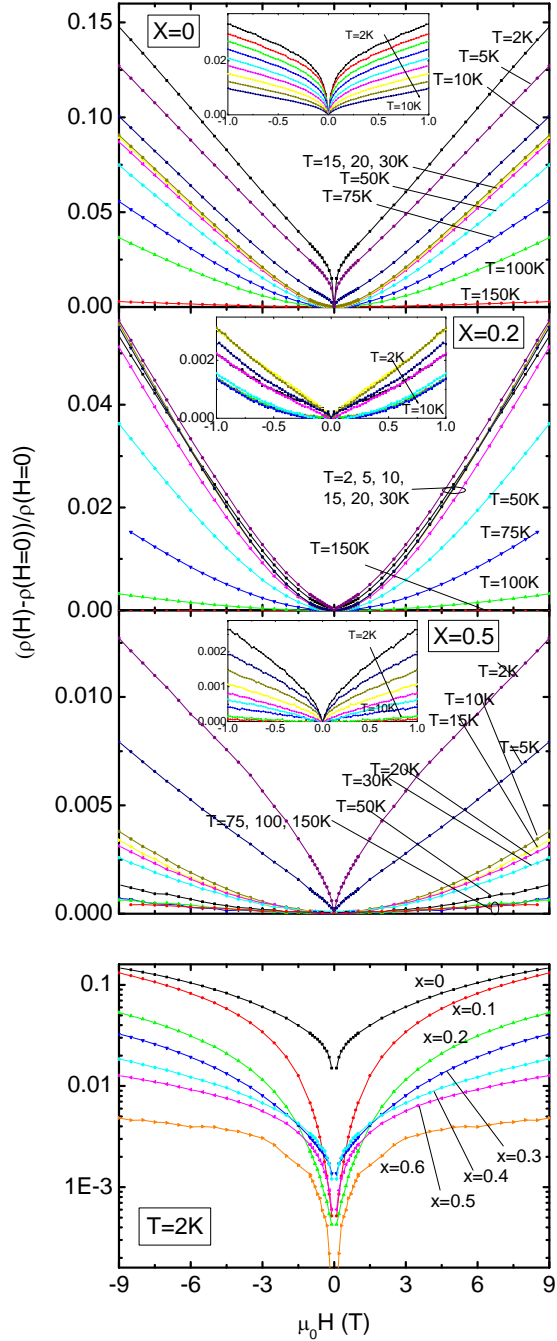

**Figure 4:** (Colour online) Magnetoresistance (ρ(H)-ρ(H=0))/ρ(H=0) curves of LaFe$_{1-x}$Ru$_x$AsO samples measured at different temperatures. For simplicity, only three samples are shown, namely x=0 in the upper panel, x=0.2 in the second panel and x=0.5 in the third panel. In the respective insets more curves measured at the lowest temperatures (T=2, 3, 4, …,10K) are shown. In the bottom panel, the magnetoresistance curves of the whole series of samples measured at T=2K are shown in logarithmic scale.

## 4. Model

In a compensated metallic system with two bands, the cyclotron theory predicts that the magnetoresistance is given by the following expression [19]:

$$\frac{\rho(H)-\rho(0)}{\rho(0)} = \frac{\sigma_1\sigma_2(\mu_1-\mu_2)^2(\mu_0 H)^2}{\sigma_1^2\left(1+\mu_2^2(\mu_0 H)^2\right)+\sigma_2^2\left(1+\mu_1^2(\mu_0 H)^2\right)+2\sigma_1\sigma_2\left(1+\mu_2\mu_1(\mu_0 H)^2\right)} \quad (1)$$

where $\sigma_i$ is the conductivity of the i-th band (i=1,2), while $\mu_i$ is its mobility, which in the above formula must be taken with the sign of the charge carriers ($\mu_i>0$ for a hole band and $\mu_i<0$ for an electron band). $\mu_0$ is the vacuum magnetic permeability. In the Drude approximation, the conductivity of each band is related to the concentration of carriers in the same band $n_i$ by $\sigma_i = q_i n_i \mu_i$, where $q_i$ are the charges of the carriers.

In turns, the Hall resistance in the same two band picture is described by:

$$R_H = \frac{1}{e} \frac{(\mu_1 \sigma_1 + \mu_2 \sigma_2) + \mu_2 \mu_1 (\mu_2 \sigma_1 + \mu_1 \sigma_2)(\mu_0 H)^2}{(\sigma_1 + \sigma_2)^2 + (\mu_2 \sigma_1 + \mu_1 \sigma_2)^2 (\mu_0 H)^2} \quad (2)$$

where $e$ is the absolute value of the electronic charge. Clearly, eq. (1) results in a positive magnetoresistance with quadratic $\propto H^2$ dependence at low fields, and a saturation to a constant value at high fields. The crossover field between these two regimes depends on the mobility values. On the other hand, in different situations a positive magnetoresistance with linear $\propto H$ dependence is expected. Among the several possible mechanisms yielding such behaviour [20, 21, 22, 23, 24], the relevant one for our case seems to be the presence of Dirac cones in the Fermi surface. Abrikosov [12,25] has predicted this linear positive contribution to magnetoresistance to appear in gapless semiconductors with linear dispersion relationship $E(\mathbf{k})$, i.e. materials with Dirac cones in their Fermi surface, under the condition that only one Landau level (LL) participates in the transport. This condition sets a lower limit to the magnetic field to be applied for this linear magnetoresistance contribution to be observed:

$$\mu_0 H > \frac{n^{2/3} \hbar}{e} \quad (3)$$

where $n$ must be intended as the net charge carrier concentration calculated only on the portions of Fermi surface forming the Dirac cones. Additionally, the LL spacing must be larger than the thermal energy $k_B T$, which sets an upper limit to the temperature:

$$T < \frac{v_F \sqrt{2\hbar e \mu_0 H}}{k_B} \quad (4)$$

The magnetoresistance contribution is then expressed by:

$$\rho = \frac{1}{2\pi} \left( \frac{e^2}{\varepsilon_\infty \hbar v_F} \right)^2 \frac{N_i}{en^2} \mu_0 H \ln(\varepsilon_\infty) \quad (5)$$

where $v_F$ is the Fermi velocity at the Dirac cones, $\varepsilon_\infty$ is the high frequency dielectric constant and $N_i$ is the impurity concentration. The situation of a Dirac cone Fermi surface is typical of graphene (where positive linear magnetoresistance is indeed observed [26]) and of other layered metals, but it has been applied also to the description of non-stoichiometric silver chalcogenides $Ag_{2+d}Se$ and $Ag_{2+d}Te$ [27], which are systems characterized by almost compensated bands as well as by disorder. Very recently, Dirac cone states have been theoretically predicted [28,29,30,31,32] in iron-based superconductors and experimentally confirmed in $Ba(FeAs)_2$ by ARPES [33,34] and transport [9,11] measurements. Before carrying out quantitative analysis of our magnetotransport data, we present some results on *ab initio* calculation on the $LaFe_{1-x}Ru_xAsO$ system, which will be used to evaluate some parameters of the system and identify possible Dirac cone features in the band structure.

**5. *Ab initio* calculations**

Band structure calculations are performed within the Density Functional Theory local spin density approximation (LSDA-DFT) in the parametrization of Perdew and Wang [35] as implemented in the all-electron LAPW code Wien2k [36]. The muffin-tin radii for La, Fe, As, and O are chosen as 2.3, 2.2, 2.0, and 1.9 Bohr radii, respectively. $R_{MT} \Box k_{max}=7$ is used as the plane-wave cutoff. We use experimental lattice parameters and internal positions shown in the inset of figure 1. The Ru-Fe pattern is simulated by an ordered structure using the $(LaFeAsO)_4$ four formula unit cell (i.e. $a=b=5.7035$Å, $c=8.755$ Å for $x=0$). We assume a antiferromagnetic stripe order as the ground state. As for the Ru doped ordered structure Ru atoms are placed along the diagonal of the 4 Fe square cell to preserve the antiferromagnetic ordering within the stripe structure. Figure 5 shows the band structure dispersion for $x$=0.00, 0.25, 0.50. Two bands (dashed and solid lines) cross the Fermi energy ($E_F$) with a third (dotted line) approaching $E_F$ at higher concentrations. The system is a compensated semimetal with equal hole and electron carrier concentrations. Holes belong to the dashed line band and electrons to the solid line one. The hole and electron bands touch, giving rise to Dirac cones that are clearly visible along the Γ−Y and Γ−X directions about 80meV below $E_F$. In figure 6 we show the Fermi surfaces (FS). Outer FS sheets are electrons and inner ones are holes. Apart from a small warping the electronic structure in these systems is nearly two-dimensional. We see that a c-axis oriented hole cylinder is present in all of the band structures at Γ. Electron FS's differ among the compositions. Two electron cylinders are present in the undoped system and four in the others. This is originated, in our ordered system, by the lowering of the system symmetry due to the substitution of Ru at Fe sites in the doped cases. This behaviour will be reflected in the actual sample by the frustration of the stripe order leading to less anisotropic band structure. Indeed, in agreement with ref. 7, we find that Ru does not own a magnetic moment in these systems. Band structures have a clear resemblance with those depicted in ref. [29].

Figure 7(a) represents the electron and hole band (solid line in figure 5) dispersion along the $k_z$=0 plane in the undoped compound. Two cone shaped structures are clearly visible. Contours lines show the crossing of the bands with $E_F$. Calculations support experimental evidence trends especially regarding the dependence of the resistivity upon doping. In figure 8 we report the density of states (DOS) for the upper valence band drawn in figure 5. We see that the DOS increases upon Ru doping. This is due to a flattening of the band structure above the Dirac points in the doped system. An evidence of this behaviour is found in the more pronounced warping of the FS in figure 6 as $x$ increases. As a consequence also carrier density increases in a non monotonic way, suggesting that Ru enhances the conduction by the increase of the DOS at $E_F$. As for the linear magnetoresistance, eq. (5) needs carriers concentration and Fermi velocity. Regarding the former, it must be pointed out that LSDA-DFT calculations notably overestimate the magnetic moment, thereby the relative positions of the bands crossing $E_F$ may not be accurate. On the other end, the value of $v_F$, that we compute directly from the expectation value of the momentum operator $\langle -i\hbar\nabla \rangle$ about the cone vertices is more robust and, moreover, our experiments do not provide an alternative and accurate way to estimate it. We therefore use our theoretical value of $v_F$. We should be conscious, however, that DFT based calculations actually overestimate the Fermi velocities. An estimate for their accuracy can be found in ref. [34], where ARPES measurements suggest a renormalization factor about 2 relative to non-magnetic standard LDA calculations. A peculiar feature of our system not present in graphene is a marked anisotropy of the in-plane component (a-b axis) of $v_F$ depending on the crystal momentum direction. Figure 7(c) shows the case of pure LaFeAsO. The velocity has its minimum value away from the Dirac point in the direction of the Brillouin zone center Γ and its maximum in the opposite direction. The ratio between maximum and minimum $v_F$ is about four. This can have effects that are not considered in Abrikosov treatment of the linear magnetoresistance. In table I we summarize some output parameters of the *ab initio* calculations.

| x. | $v_F$ (Km/s) | $n_e$, $n_h$ (e/cell) |
|---|---|---|
| 0.0 | 400-100 | 0.044 |
| 0.25 | 250-60 | 0.131 |
| 0.50 | 200-40 | 0.125 |

**Table I:** Output parameters of the *ab initio* calculations, namely anisotropic Fermi velocities with their range of variability along the different directions and carrier densities ($n_e$=$n_h$).

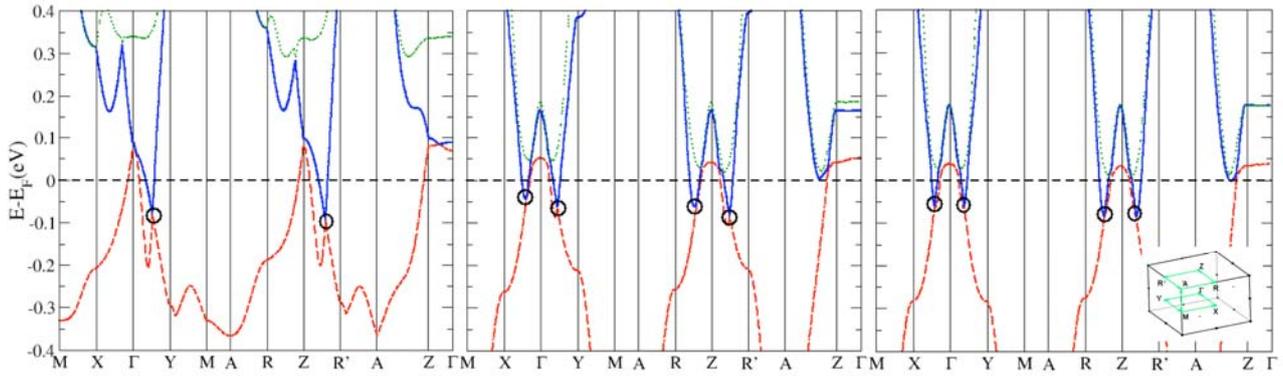

**Figure 5:** (Colour online) Band structure dispersion along the high symmetry directions of the Brillouin zone (see inset) for the tetragonal four formula unit cell of LaRu$_x$Fe$_{1-x}$AsO for $x$=0 (left), 0.25 (middle), 0.5 (right). Cones are marked by circles. We show only selected bands crossing the Fermi energy.

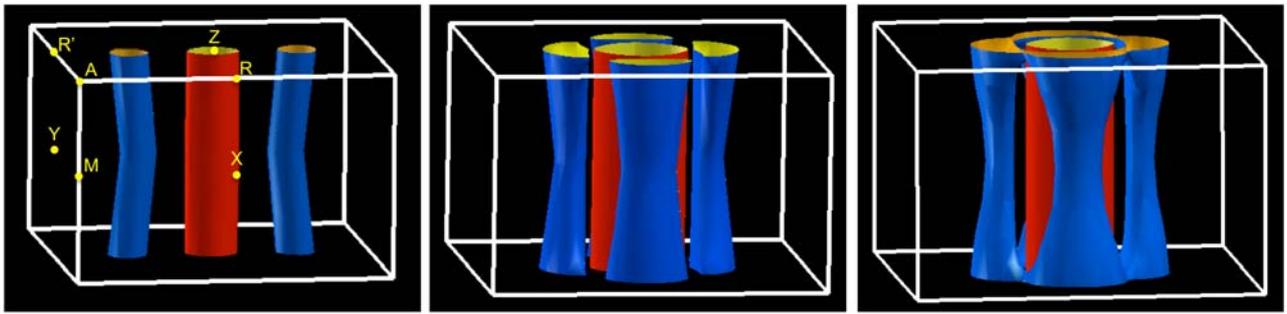

**Figure 6:** (Colour online) Fermi surfaces of LaRu$_x$Fe$_{1-x}$AsO for $x$=0 (left), 0.25 (middle), 0.5 (right).

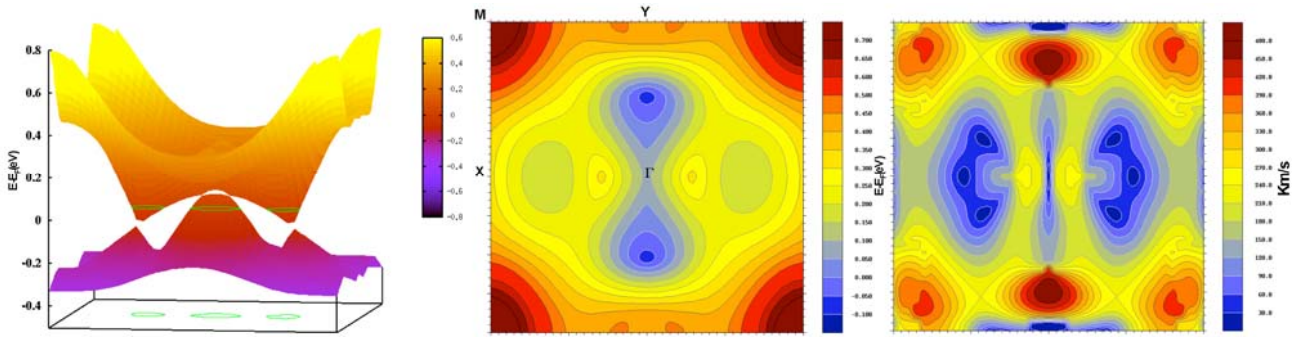

**Figure 7:** (Colour online) Band dispersion and group velocity in LaFeAsO . Panel (a) shows the band energy along the $k_z$=0 plane. Contours highlight the Fermi level crossing. Panel (b) electron band, contours evidence the anisotropy of the cone shape. Panel (c) electrons group velocity.

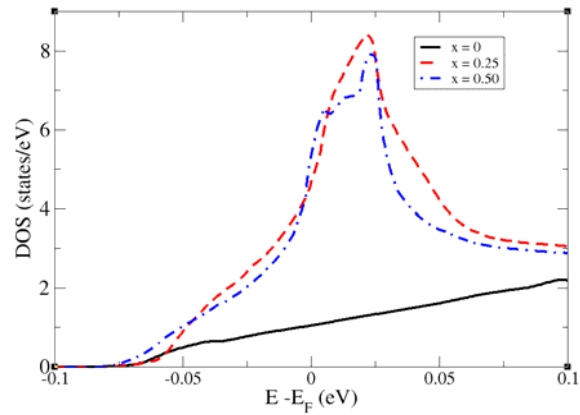

**Figure 8:** (Colour online) Density of states for the lower conduction band (solid line in figure 5) as a function of Fermi level, for different Ru dopings.

## 6. Data analysis
*Cyclotron magnetoresistance*

We try to fit quantitatively our magnetoresistance data, as a sum of two terms of the type of eq. (1) and (5), namely:

$$\frac{\rho(B)-\rho(0)}{\rho(0)} = \frac{\alpha(\mu_0 H)^2}{\beta+(\mu_0 H)^2} + \gamma|\mu_0 H| \qquad (6)$$

with α, β and γ as fitting parameters. In figure 9, two examples of fit are displayed, for the x=0 sample at temperatures 2K and 50K. In the insets, the linear and cyclotron contributions are shown separately. In order to get the carrier concentrations and mobilities from α and β, we need some further constraints and considerations. Two constraints could be the experimental values of Hall resistance and resistivity. Moreover, we observe that by varying the carrier concentrations throughout a very wide range of realistic values, the mobilities turn out to be of the same sign for most of temperatures and Ru concentrations (except for x=0.3 and x=0.5 at T≥75K, and also x=0.6 at all temperature, where also fits with mobilities of opposite sign are possible), which means that two bands of the same type of carriers dominate the transport. This is particularly evident at low temperature, where the cyclotron magnetoresistance term has a very low saturation value $\frac{\rho(B)-\rho(0)}{\rho(0)} \approx 0.02$, but at the same time it saturates already at magnetic fields as low as 0.5T, causing the characteristic sharp dip to appear in the experimental curves. This behaviour is possible only by assuming two very large mobilities of the same sign, which almost cancel out in eq (1). From the negative values of Hall resistance data in figure 3 we then assume that the two bands are electron type. Recently a large difference between the scattering rates of holes and electrons for scattering by spin and charge fluctuations has been predicted in the parent compounds of Fe-based pnictides [37]. As a consequence of this difference, the transport properties should be dominated by small parts of the electron pockets where the lifetimes are long and the Fermi velocities are high. Indeed, by comparison with the theoretical band structure of figure 5 and consistently with this theoretical prediction, we identify these two electron channels with two directions of the anisotropic electron Dirac cone centred along the Γ−Y and Γ−X directions, which behave as two channels in parallel in polycrystalline samples. As for the carrier concentration of this electron band, our *ab initio* calculations predict values equal to 0.044, 0.131 and 0.125 per unit cell for x=0, 0.25 and 0.5, respectively, as reported in table I. By fixing these parameters, the $R_H$ values turn out to be smaller than the experimental ones by one order of magnitude and the resistivities smaller than the experimental ones by two-three orders of magnitude. This indicates that most of the carriers are condensed in the SDW state and only a small fraction takes part in the transport. We therefore assume as free parameter the carrier concentration ($n_1=n_2$, as it is the same anisotropic band), together with the mobilities $\mu_1$ and $\mu_2$, forcing the fit to reproduce exactly the experimental magnetoresistance curves and the $R_H$ values, while the experimental resistivity value is not used as a constraint. In figure 10 the two band mobilities (left-hand axes) and carrier concentrations (right-hand axes) resulting from the fitting of all samples at different temperatures are shown. The same vertical scales are maintained in all the panels for easier comparison. The carrier densities clearly show the typical condensation behaviour below the $T_{SDW}$ temperatures reported in figure 3, marked by vertical lines. Postponing the discussion on the low temperature data (T<15K), for which we will see that the consistency between the in-field transport parameters (fitting values of mobility) and the zero field ones (resistivity) is not necessarily required, the experimental resistivities are also reproduced within the same order of magnitude as long as the carrier densities are much smaller than the theoretical values, obtained by linear interpolation of the values in table I and indicated as horizontal dashed lines (more specifically, the experimental resistivities are reproduced when approximately $n_1=n_2<0.1$ electrons per unit cell). In this regime the mobilities are plotted as filled symbols in figure 10. For large x and for temperatures approaching $T_{SDW}$ values, the fitting carrier concentrations approach the theoretical values. In this regime, it is likely that also a significant portion of holes contribute to transport, so that a fit with hole and electron bands is necessary to account for the experimental resistivity values. In this case, the fit parameters are not anymore determined univocally, but the qualitative trend of mobilities is the same as that resulting from the fits with two electron bands. Hence, in figure 10, the results of fits with two electron bands are displayed also in this regime, but as open symbols, as a reminder for the reader to take the quantitative values with caution.

In the bottom right-hand panel of figure 10, we plot the highest values of mobility at 5K and the lower values at 50K versus the Ru content. For comparison, we also plot the Hall mobilities $\mu_H$ at the same temperatures, extracted as the ratio of experimental Hall resistance to resistivity $R_H/\rho$, in a single band approximation. It can be seen that the $\mu_H$ values are more than three orders of magnitude smaller than the mobility values extracted by fitting the magnetoresistance curves. This discrepancy is due in part to the inadequacy of the single band description, but the main reason is that the in-field and zero-field transport mechanisms are intrinsically different in these systems, as it will be explained in the following.

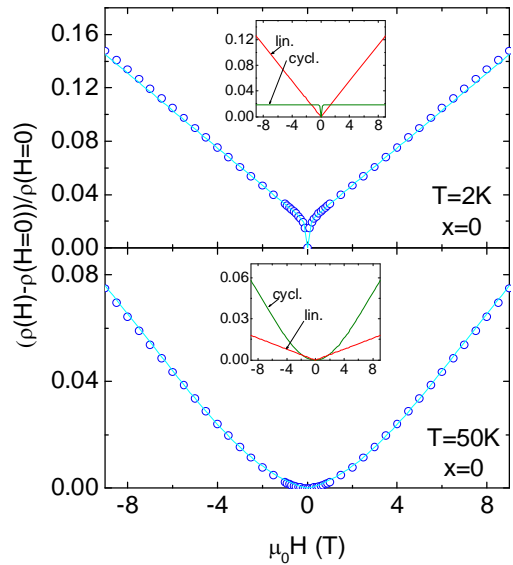

**Figure 9:** (Colour online) Experimental magnetoresistance $(\rho(H)-\rho(H=0))/\rho(H=0)$ curves of the LaFeAsO sample measured at 2K and 50K (open symbols) and corresponding fitting curves by eq. (6). In the insets, the linear and cyclotronic contributions are shown separately.

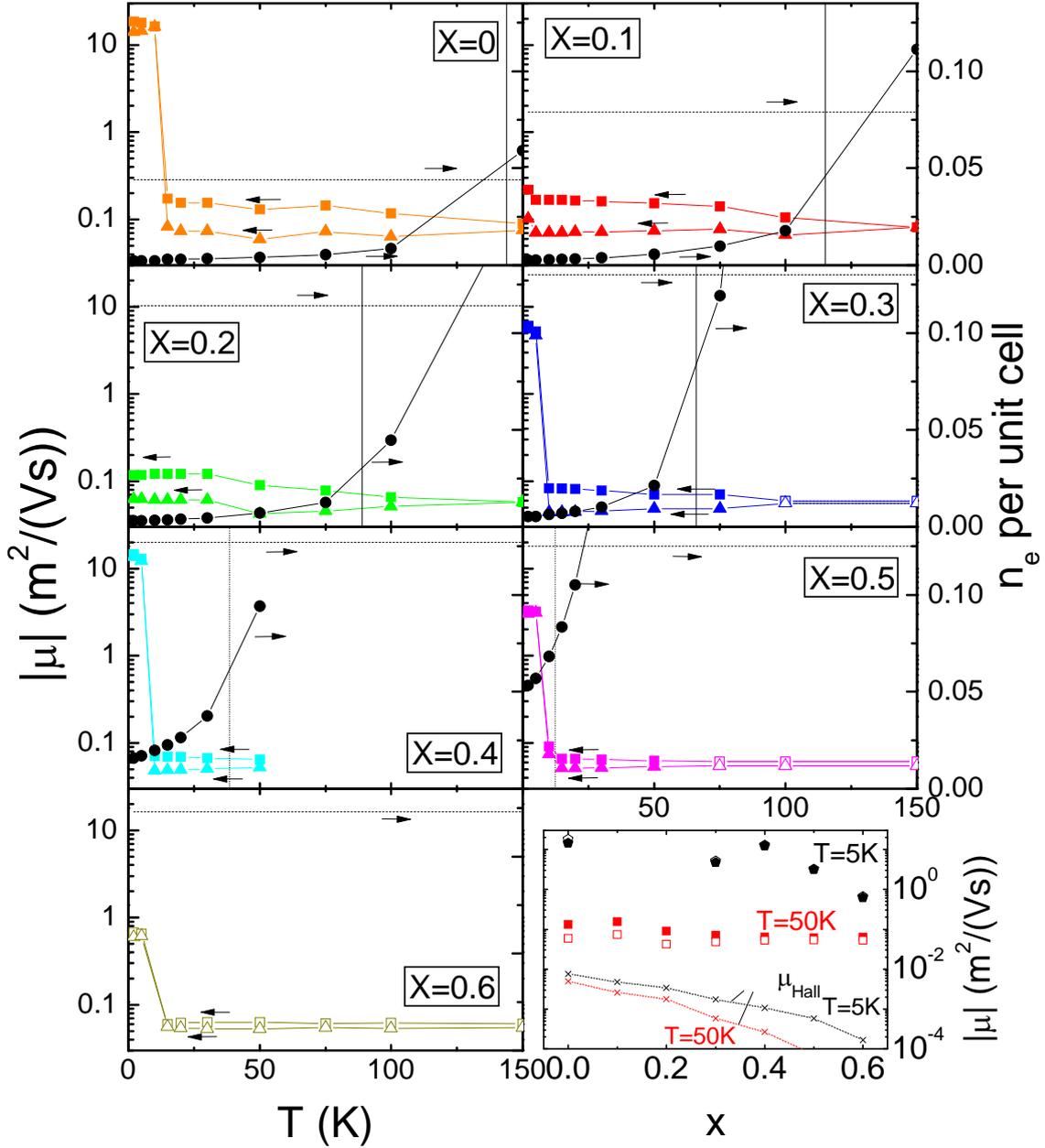

**Figure 10:** (Colour online) Mobilities of the electron bands as a function of temperature, extracted by fitting magnetoresistance curves of the LaFe$_{1-x}$Ru$_x$AsO samples with a two-band model (filled squares and triangles, left-hand axes); the open symbols indicate the mobilities extracted in a regime where also holes take part in the transport, and therefore the fit are less reliable (open squares and triangles, left-hand axes). Also plotted are the carrier concentrations of the electron bands as a function of temperature, extracted from the same fits (filled circles, right-hand axes). The theoretical values of carrier concentrations are indicated as horizontal dashed lines (right-hand axes). The measured T$_{SDW}$ temperatures are indicated as vertical continuous lines, while the extrapolated T$_{SDW}$ values are indicated as vertical dotted lines. The same vertical scales are maintained in all the 7 panels, for better comparison. In the x=0.6 panel, the carrier concentrations are slightly above of the displayed scale. In the bottom left panel, the high mobilities at T=5K and those at 50K are plotted as a function of the Ru content x. For comparison, in the same panel also the experimental values of the Hall mobilities are plotted.

In the other seven panels of figure 10, it can be seen that the behaviour of mobility curves is qualitatively similar for the whole series of samples. The most striking feature which emerges from the temperature dependencies is the abrupt rise in mobility that occurs below 10K-15K. For example, in the x=0 sample the mobility increases by a factor 200 within a few K. Moreover, the low temperature mobility values are strikingly large, namely for the x=0 sample we extract a value of 18.6 m²/(Vs), which is in the typical range of high mobility semiconductors. The low temperature mobility and the related drop above 10-15K seems to be pretty sensitive to the Ru content and non monotonic as a function of x. Indeed, it is observed in the x=0 and 0.3≤x≤0.6 samples, but not in the x=0.1 and x=0.2 ones. In particular, the magnetoresistance dip at low fields is the signature of the huge mobility values below 15K, as the dip comes out just from the cyclotron magnetoresistance term, which is quickly saturating as a consequence of the huge mobility (see upper panel of figure 9). On the contrary, the values of mobility above 15K decrease weakly and monotonically as a function of x. This is visible in the lower right-hand panel of figure 10, where the mobilities are plotted versus the Ru content x, at two different temperatures T=5K and T=50K. The drop of mobility at 10-15K is not correlated to the SDW, as the SDW temperature is much larger and it is significantly suppressed by Ru substitution, while the temperature of the mobility rise remains rather unaffected. It is not related to the disappearance of the Dirac cones either, indeed the linear magnetoresistance term, signature of Dirac cones, persists up to much larger temperatures (see figure 11 and following description). We attribute these exceptionally high values of the low temperature mobilities to the first LL in the Dirac cones, whose dispersion relationship is associated to a vanishingly small effective mass and hence a huge mobility, despite the presence of disorder and localization. However, when the temperature rises above 15K, corresponding to 1.3meV, which is the energy scale of the LL spacing $\Delta = v_F \sqrt{2\hbar e \mu_0 H}$, the second LL with related inter-LL scattering and with its larger effective mass become involved in the transport. Hence the abrupt change in mobility at 15K is mainly due to the fact that when the first LL is fully occupied and the second one is empty, if the LL spacing is smaller than the thermal energy there are no available empty final states for the scattered carriers and scattering is effectively inhibited.

The x=0.1 and x=0.2 samples do not exhibit the huge mobility values related to the first LL in the Dirac cones. One possibility is that these samples are affected by a high level of crystallographic disorder which broadens the quantum levels. It has been indeed observed that the mobility of $SmFe_{1-x}Ru_xAsO_{0.85}F_{0.15}$ compounds does not decrease monotonically with x, but decreases down to x=0.25 and then increases again, suggesting that a local order is re-established around x=0.5. However, looking at the resistivity data in figure 3 as well as at the mobility data at 50 K in figure 10, the x=0.1 and x=0.2 samples are in trend with the others and do not suggest that these samples suffer a larger level of disorder. Another possible explanation is that the Fermi level does not shift monotonically with increasing Ru. It can be assumed that the Fermi level initially increases with x, so that for x=0.1-0.2 the second LL is filled, while for further increase in x the Fermi level decreases again, leaving again a single LL occupied. This interpretation is qualitatively supported by *ab initio* calculations that predict such non monotonic behaviour of the electron and hole carrier density with increasing Ru content, as reported in table I, namely it increases from x=0 to x=0.25 and decreases from x=0.25 to x=0.5.

It is clearly seen that this abrupt rise of mobility has no counterpart in other transport data, namely resistivity and $R_H$, which behave smoothly in this temperature range (see figure 3). Indeed, in this high mobility regime (T<15K) the resistivities calculated from the fitting parameters as $\rho = \left( \sum_i q_i \mu_i n_i \right)^{-1}$ are almost two orders of magnitude smaller than the experimental values. This discrepancy is reconciled by considering that magnetoresistance probes a different mean free path from that probed by resistivity, or, in other words, that the in-field transport is intrinsically different from the zero field transport. Namely, this occurs because of two reasons, namely that in a magnetic field (i) LLs form in the density of states and (ii) charge carriers follow cyclotron orbits. Indeed, in a large enough magnetic field, the formation of LLs marks the onset of a substantially different transport regime, with enhanced mobility and mean free path. The zero field mean free path λ could be in principle calculated as $\lambda = \frac{\mu v_F m_{eff}}{e}$, using for $m_{eff}$ the expression for the effective mass of Dirac electrons in zero field (see appendix):

$$m_{eff} = \sqrt{\frac{\pi \hbar^2 n_{2D}}{v_F^2}} \tag{7}$$

where $n_{2D}$ is the sheet carrier density. Unfortunately, the zero field mobility cannot be extracted from our experimental data and the Hall mobility extracted in a single band approximation is unreliable. If we used for µ and $n_{2D}$ the parameters extracted from the magnetoresistance fits, which are valid only in a finite magnetic field, we would get only an upper limit for the zero field mean free path, as the actual zero field mobility is certainly much lower that the in-field one and moreover, it may be limited by domain boundaries in our samples. Indeed, the average domain size measured by polarized optical microscopy is in the 0.1-0.2 µm range (see figure 2). On the contrary, in a finite magnetic field, the mean free path λ can be estimated from our transport data in two ways. (i) For the x=0 sample at 2K, we observe a saturation of magnetoresistance at µ₀H=0.5T (see inset in the upper panel of figure 9); assuming that saturation occurs when the mean free path is equal to a cyclotron circumference, we get a λ≈0.5µm. (ii) λ can be also evaluated using the expression of the effective mass in Dirac cone electrons in the ν-th LL (see appendix):

$$m_{eff(\nu)} = \sqrt{\frac{2e\hbar\mu_0 H\nu}{v_F^2}} \tag{8}$$

and then the expression $\lambda = \frac{\mu v_F m_{eff(\nu)}}{e}$. For the x=0 sample, below 15K, with only the lowest LL occupied (ν=1), we get λ≈0.3µm. The similarity of these two independent evaluations confirms the consistency of our analysis. We also point out that, at the lowest temperatures, the in-field λ value is in the sub-micron range, which is comparable with the domain size, but it is not limited by domain boundaries as long as electrons are deflected into cyclotron orbits. Indeed, in a semiclassical description, Dirac electrons of the lowest LL are confined to move along cyclotron orbits whose radius is given by (see appendix):

$$R_{cycl} = \frac{v_F}{\omega_c} = \sqrt{2\hbar / e\mu_0 H} \tag{9}$$

where $\omega_c$ is the cyclotron frequency. For example, already at fields as low as µ₀H=0.5T, we get a $R_{cycl}$ value as low as 0.05µm, which is smaller than the average domain size in the 0.1-0.2 µm range.

*Linear magnetoresistance*
From the magnetoresistance fit, we also extract the behaviour of the linear magnetoresistance term described by eq. (5). In figure 11 we plot the best fit values of the linear magnetoresistance coefficient γ as a function of temperature at different Ru contents. As expected, γ decreases with increasing temperature, but it is still non vanishing at temperatures as high as 100K for the x=0 sample. Indeed, from eq. (4), assuming the average calculated Fermi velocity $v_F$≈100 Km/s, we get a maximum temperature for the observation of this effect around 125 K. As for the magnetic field limit, eq. (3) overestimates the minimum necessary field, which turns out to be 20T, while in our experiment a few Tesla are already enough. From the inset of figure 11 it is seen that the coefficient γ decreases with increasing Ru content. We can compare these data with eq. (5). From *ab initio* calculation we assume that the Fermi velocity does not vary appreciably with the Ru content (see table I). We estimate $N_i$ from the λ as $N_i \approx \left(\frac{4}{3}\pi\left(\frac{\lambda}{2}\right)^3\right)^{-1}$, which turn out to be much lower than the Ru concentration and thus not directly related to it. Finally, we take carrier concentration values from the fits. The expected values of $\gamma \propto \dfrac{N_i}{v_F^2 n^2}$ from eq. (5) reproduce qualitatively the behaviour with temperature and Ru content, given the dependence of $n$ on temperature and Ru reported in figure 10. Moreover, eq. (5) gives the correct order of magnitude for $\gamma$ for $x \leq 0.2$ above 20K, which is quite satisfying given the uncertainty on the carrier and impurity concentrations.

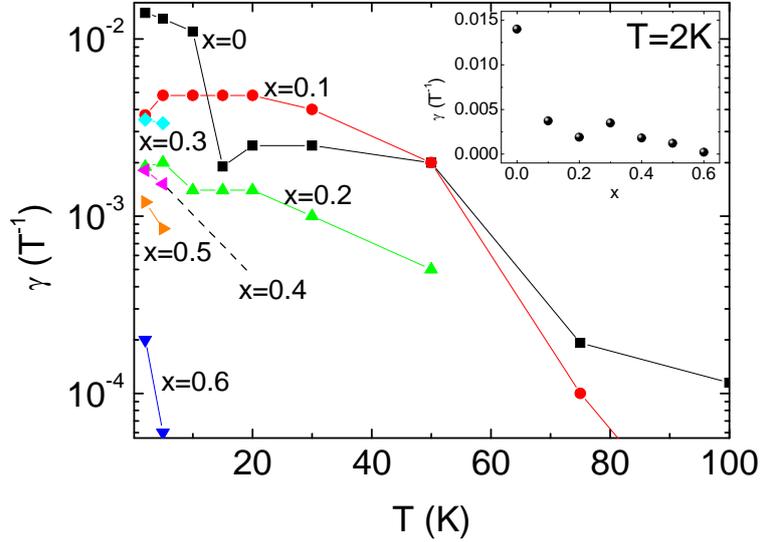

**Figure 11:** (Colour online) Coefficient of the linear term of magnetoresistance extracted by fitting the experimental data of LaFe$_{1-x}$Ru$_x$AsO samples, plotted as a function of temperature. In the inset, the data at T=2K are plotted as a function of the Ru content x.

## 7. Discussion
*Relationship between Dirac cones and SDW ordering*
*Ab initio* calculations suggest that features related to Dirac cones can be observed only below $T_{SDW}$, as the antiferromagnetic ordering induces a modulation which doubles the cell periodicity and gives actual physical significance to Dirac cones. On the other hand, the gap associated to the SDW ordering itself may open at the Dirac cone vertices, thus destroying Dirac cones. Our series of samples are suitable for the exploration of this relationship between Dirac cones and SDW ordering, in that $T_{SDW}$ is gradually suppressed by Ru substitution and the corresponding trend of Dirac cones related features, such as the linear magnetoresistance, can be inspected. We can compare $T_{SDW}$ values defined by the $d\rho/dT$ maximum - or alternatively $T^*_{SDW}$ values defined as the temperatures at which the carrier concentrations extracted from the fits are reduced to 50% of the theoretical values - with the maximum temperature at which the linear magnetoresistance is visible $T_\gamma$, taken from the plots of figure 11. The results are plotted in figure 12 as a function of the Ru content. Clearly there is a correlation between these temperatures, indicating that the SDW order is not detrimental to Dirac cones, as the SDW gap does not open at the Dirac cones. Moreover, the correlation between $T_{SDW}$ and $T_\gamma$ may be also due to the fact that as the carriers condense, the quantum limit is approached and linear magnetoresistance related to Dirac cones appears. This also explains why Dirac cones are observed up to larger $T_\gamma$ temperatures with decreasing x, despite the condition of eq. (4) is almost the same for all x values.

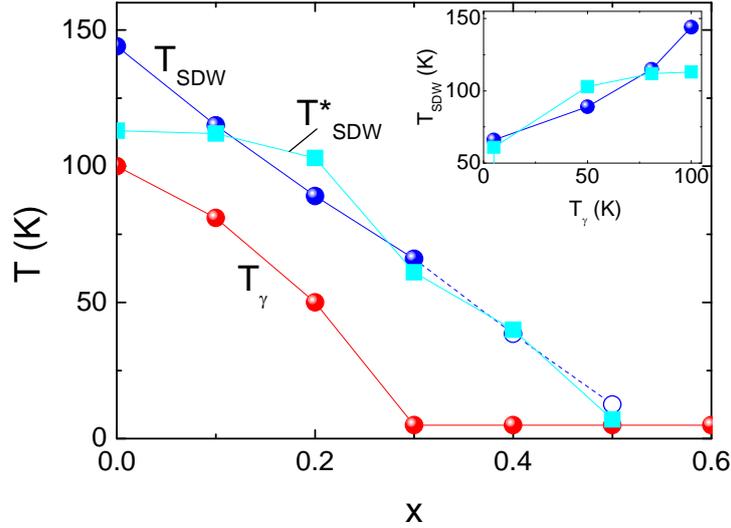

**Figure 12:** (Colour online) $T_{SDW}$, defined from SDW feature in the resistivity curves, $T^*_{SDW}$, defined as the temperatures at which the carrier concentrations extracted from the fits are reduced to 50% of the theoretical values and $T_\gamma$, defined as the maximum temperatures at which the linear magnetoresistance is visible, as a function of Ru content. Open symbols indicate extrapolated values. In the inset $T_{SDW}$ and $T^*_{SDW}$ are plotted versus $T_\gamma$, in order to emphasize the correlation.

*Relationship between Dirac cones and superconductivity*
One further remark must be made about the relationship between Dirac cones and superconductivity. A crucial issue is to understand whether there exists a direct relationship between Dirac cones and superconductivity or else an indirect one, via their respective relationship with the SDW ordering. Tanabe et al. [11] has found that Dirac cones, present in the 122 parent compound, survive also in doped compounds, where they coexist with superconductivity. An opposite result has been found in $Eu(Fe_{1-x}Co_x)_2As_2$ by Matusiak et al. [38], where Co doping yields suppression of SDW, appearance of superconductivity and vanishing Dirac cone behaviour at the same time.
In the 1111 family, the linear magnetoresistance term, well visible in the parent compounds, disappears in the superconducting compounds. For example, the linear magnetoresistance in SmFeAsO (see figure 2 in ref. [39]) is not seen anymore in the $SmFe_{1-x}Ru_xAsO_{0.85}F_{0.15}$ series (see figure 11 in ref. [7]). The lack of Dirac cone signature in the above 1111 superconducting compounds may have a twofold explanation: (i) the absence of SDW carrier condensation places the Fermi level well above the Dirac cone vertex and the quantum limit of one LL occupied is not anymore approached in superconducting sample; (ii) the absence of SDW antiferromagnetic modulation reduces the lattice periodicity, making Dirac cones artifact features of band folding, with no counterpart in experimental phenomenology. The experimental data cannot sort out this puzzle.

*Relationship between Dirac cones and disorder*
In our $LaFe_{1-x}Ru_xAsO$ samples, the isoelectronic Ru substitution, especially at low Ru contents $x \leq 0.5$ [7], allows us to explore the robustness of the Dirac cones against disorder. The Dirac cone states in Fe-based superconductors are bulky features induced by a three-dimensional zone folding of bands related to nodes which survive the gap opening in the SDW state. Thus, they are different from the Dirac cones of other systems such as the two-dimensional graphene and the space-inversion symmetry broken surface of topological insulators, while they are rather more akin to the Dirac cones of bismuth. Dirac states enjoy different degrees of protection against disorder and interactions in different systems. For example, in silver chalcogenides such as $Ag_{2+d}Se$ and $Ag_{2+d}Te$ [27] it has been found that the presence of disorder is crucial for the appearance of the linear magnetoresistance term. Indeed, these systems are narrow bandgap semiconductors which are turned into gapless semiconductors by the band edge tailing related to disorder, where Dirac cones appears at the points of the reciprocal space where conduction band and valence band touch. In other Dirac cone systems such as graphene, disorder has been predicted to broaden the quantum levels close to the Dirac point, yielding strong modifications in the physical properties [40]. On the other hand, some experimental data indicate that the mobility is almost unchanged upon introduction of disorder [41].

In the case of LaFe$_{1-x}$Ru$_x$AsO samples, the linear magnetoresistance term is well visible in the Ru free sample and it decreases weakly with increasing Ru content. In section 6.2 it is shown that the estimated defect concentration $N_i$ seems not to be correlated with the Ru concentration, indicating that Ru impurities are not effective scatterers for Dirac electrons. The decreasing trend of the coefficient γ as a function of x, shown in the inset of figure 11, is thus possibly related to the downshift of the T$_{SDW}$ with increasing x, which implies a lower carrier condensation at a given temperature and thus a lower $\gamma \propto n^{-2}$. Hence the Ru content cannot be considered a good measure of the disorder met by Dirac electrons. It is interesting to note that, on the contrary, the resistivity upturn is largely influenced by the Ru substitution (see figure 3), indicating that it yields significant structural disorder. Moreover, in a homologous series of SmFe$_{1-x}$Ru$_x$AsO$_{0.85}$F$_{0.15}$ samples [7] it has been found that Ru substitution has a pair-breaking effect of T$_c$. This suggests that the mechanisms involving Dirac electrons and Cooper pairs are not closely related.

## 8. Conclusions

We use a combined experimental and theoretical approach to investigate the band structure and magnetotransport properties of the LaFe$_{1-x}$Ru$_x$AsO system (x form 0 to 0.6), in order to get information on the parent compound of the so called 1111 family of Fe-based superconductors. *Ab initio* calculations indicate the presence of Dirac cones with anisotropic Fermi velocity along the Γ−Y and Γ−X directions of the Brillouin zone. This finding is supported by the observation of a linear term in the magnetoresistance curves, which survives up to temperatures as high as 100K and Ru doping up to x=0.6. Indeed, despite the fact that disorder is effectively introduced into the system by Ru substitution, its effect in the band structure at the Dirac cones is pretty weak.

A quantitative fit of magnetoresistance curves allows to extract the band mobilities. It turns out that transport is dominated by electron bands, in agreement with theoretical predictions of scattering rates by spin and charge fluctuations, and the corresponding mobilities rise abruptly below 15K, reaching values as large as 18.6 m$^2$/(Vs) at 2K in the LaFeAsO sample. The origin of such abrupt rise is related to the extreme quantum limit of occupation of a single Landau level in the Dirac cones. These huge mobility values are the first clear evidence of outstanding transport properties of Dirac cones in Fe-based pnictides.

We find a clear correlation between the temperature ranges of linear magnetoresistance, associated to Dirac cones, and SDW carrier condensation, also supported by the theoretical predictions that the SDW gap does not open at the Dirac cones.

Finally, no evidence of direct relationship between Dirac cones and superconductivity comes out from our analysis.

## 9. Appendix

In this section we derive some useful expressions used in the paper, namely (i) the cyclotron radius for the lowest LL of Dirac cones (eq. (9)), (ii) the effective mass of Dirac electrons in zero field (eq. (7)) and (iii) the effective mass of Dirac electrons in the ν-th LL (eq. (8)).

i) Given the linear dispersion relation characteristic of Dirac cones, the Fermi energy is written as:
$$E_F = \hbar k_F v_F \qquad (a.1)$$
while the energy of the lowest LL is given by:
$$E_{\nu=1} = v_F \sqrt{2\hbar e \mu_0 H} \qquad (a.2)$$

Assuming that only the lowest LL is occupied we fix $E_F = E_{v=1}$ and get an expression for the Fermi wavevector:

$$k_F = \sqrt{\frac{2e\mu_0 H}{\hbar}} \quad (a.3)$$

The cyclotron frequency is defined as:

$$\omega_c = \frac{1}{k_F}\frac{dk_F}{dt} \quad (a.4)$$

Using the expression for Lorentz force $F_L = \frac{dk_F}{dt} = ev_F\mu_0 H$ and eq. (a.3), we can write eq. (a.4) as:

$$\omega_c = v_F\sqrt{\frac{e\mu_0 H}{2\hbar}} \quad (a.5)$$

The cyclotron radius is finally obtained as:

$$R_{cycl} = \frac{v_F}{\omega_c} = \sqrt{\frac{2\hbar}{e\mu_0 H}}$$

which is eq. (9).

ii) From the definition of the cyclotron frequency and from the dispersion relation eq. (a.1), we write:

$$m_{eff} = \frac{e\mu_0 H}{\omega_c} = \frac{\sqrt{2\hbar e\mu_0 H}}{v_F} = \frac{E_F}{v_F^2} \quad (a.6)$$

The Fermi wavevector in two dimensions is determined by the sheet carrier density $n_{2D}$ per Dirac cone; in this case we have two Dirac cones $n_{DC}=2$ (see two electron cylinders in the left panel of figure 6), hence:

$$k_F = \sqrt{\frac{2\pi n_{2D}}{n_{DC}}} = \sqrt{\pi n_{2D}} \quad (a.7)$$

From eq. (a.1), (a.6) and (a.7), the effective mass of Dirac electrons in zero magnetic field is easily obtained:

$$m_{eff} = \sqrt{\frac{\pi\hbar^2 n_{2D}}{v_F^2}}$$

which is eq. (7).

iii) The energy of the $v$-th Landau level is:

$$E_v = v_F\sqrt{2\hbar e\mu_0 H v} \quad (a.8)$$

We can substitute eq. (a.8) into (a.6) and then assume that the Fermi energy coincides with the $v$-th Landau level. We thus obtain:

$$m_{eff(v)} = \frac{E_v}{v_F^2} = \sqrt{\frac{2e\hbar\mu_0 H v}{v_F^2}}$$

which is eq. (8).

**Acknowledgements**
This work was partially supported by Grant No. PRIN2008XWLWF9.
F.B. acknowledges support from CASPUR under the Standard HPC Grant 2011 and CINECA Project class C no. HP10CYKFIW.

The authors are grateful to Gianrico Lamura for scientific discussion.

**References**

[1] Y.Kamihara, T.Watanabe, M.Hirano, H.Hosono, J. Am. Chem. Soc. 130, 3296 (2008)
[2] C.Wang, L.Li, S.Chi, Z.Zhu, Z.Ren, Y.Li, Y.Wang, X.Lin, Y.Luo, S.Jiang, X.Xu, G.Cao, Z.Xu, Europhys. Lett. 83, 67006 (2008)
[3] I.I.Mazin, D.J.Singh, M.D.Johannes, M.H.Du, Phys. Rev. Lett. 101, 057003 (2008)
[4] D.H.Lu, M.Yi, S.-K.Mo, A.S.Erickson, J.Analytis, J.-H.Chu, D.J.Singh, Z.Hussain, T.H.Geballe, I.R.Fisher, Z.-X.Shen, Nature 455, 81 (2008)
[5] F.Wang, H.Zhai, D.-H.Lee, Phys. Rev. B 81 184512 (2010)
[6] B.J.Arnold, S.Kasahara, A.I.Coldea, T.Terashima, Y.Matsuda, T.Shibauchi, A.Carrington, Phys. Rev. B 83, 220504(R) (2011)
[7] M.Tropeano, M.R.Cimberle, C.Ferdeghini, G.Lamura, A.Martinelli, A.Palenzona, I.Pallecchi, A.Sala, I.Sheikin, F.Bernardini, M.Monni, S.Massidda, M.Putti, Phys. Rev. B 81, 184504 (2010)
[8] M.A.McGuire, D.J.Singh, A.S.Sefat, B.C.Sales, D.Mandrus, J. Solid State Chem. 182, 2326 (2009)
[9] K.K.Huynh, Y.Tanabe, K.Tanigaki, Phys. Rev. Lett. 106, 217004 (2011)
[10] D.Bhoi, P.Mandal, P.Choudhury, S.Pandya, V.Ganesan, Appl. Phys. Lett. 98, 172105 (2011)
[11] Y.Tanabe, K.K.Huynh, S.Heguri, G.Mu, T.Urata, J.Xu, R.Nouchi, N.Mitoma, K.Tanigaki, Phys. Rev. B 84, 100508(R) (2011)
[12] A.A.Abrikosov, Europhys. Lett. 49, 789 (2000)
[13] D.Hsieh, D.Qian, L.Wray, Y.Xia, Y.S.Hor, R.J.Cava, M.Z.Hasan, Nature 452, 970 (2008)
[14] Y.Xia, D.Qian, D.Hsieh, L.Wray, A.Pal, H.Lin, A.Bansil, D.Grauer, Y.S.Hor, R.J.Cava, M.Z.Hasan, Nature Phys. 5, 398 (2009)
[15] A.V.Geim, K.S.Novoselov, Nature Mater. 6, 183 (2007)
[16] L.Li, J.G.Checkelsky, Y.S.Hor, C.Uher, A.F.Hebard, R.J.Cava, N.P.Ong, Science 321, 547 (2008)
[17] A.Damascelli, Z.Hussain, Z.X.Shen, Rev. Mod. Phys. 75, 473 (2003)
[18] P.Bonfà, P.Carretta, S.Sanna, G.Lamura, G.Prando, R.De Renzi, A.Palenzona, M.Tropeano, A.Martinelli, M.Putti, manuscript in preparation
[19] J.M.Ziman, *Principles of the Theory of Solids*, Cambridge University Press, Cambridge (1972)
[20] F.J.Ohkawa, Phys. Rev. Lett. 64, 2300 (1990)
[21] O.Entin-Wohlman, Y.Levinson, A.G.Aronov, Phys. Rev. B 49, 5165 (1994)
[22] C.Herring, J. Appl. Phys. 31, 1939 (1960)
[23] J.Fenton, A.J.Schofield, Phys. Rev. Lett. 95, 247201 (2005)
[24] M.P.Delmo, S.Yamamoto, S.Kasai, T.Ono, K.Kobayashi, Nature 457, 1112 (2009)
[25] A.A.Abrikosov, Phys. Rev. B 58, 2788 (1998)
[26] S.Cho, M.S.Fuhrer, Phys. Rev. B 77, 081402 (2008)
[27] M.Lee, T.F.Rosenbaum, M.L.Saboungi, H.S.Schnyders, Phys. Rev. Lett. 88, 066602 (2002); H.S.Schnyders, M.L.Saboungi, T.F.Rosenbaum, Appl. Phys. Lett. 76, 1710 (2000)
[28] H.Fukuyama, JPSJ Online-News and Comments, 77 (May 12, 2008)
[29] Y.Ran, F.Wang, H.Zhai, A.Vishwanath, D.-H.Lee., Phys. Rev. B 79, 014505 (2009)
[30] T.Morinari, E.Kaneshita, T.Tohyama, Phys. Rev. Lett. 105, 037203 (2010)
[31] M.Z.Hasan, B.A.Bernevig, Physics 3, 27 (2010)
[32] Z.P.Yin, K.Haule, G.Kotliar, Nature Physics 7, 294 (2011)
[33] Y.Kim, H.Oh, C.Kim, D.Song, W.Jung, B.Kim, H.J.Choi, C.Kim, B.Lee, S.Khim, H.Kim, K.Kim, J.Hong, Y.Kwon, Phys. Rev. B 83, 064509 (2011)
[34] P.Richard, K.Nakayama, T.Sato, M.Neupane, Y.-M.Xu, J.H.Bowen, G.F.Chen, J.L.Luo, N.L.Wang, X.Dai, Z.Fang, H.Ding, T.Takahashi, Phys. Rev. Lett. 104, 137001 (2010)
[35] J.P.Perdew, Y.Wang, Phys. Rev. B 45, 13244 (1992)
[36] P.Blaha, K.Schwarz, G.K.H.Madsen, D.Kvasnicka, J.Luitz, *WIEN2k: An Augmented Plane Wave Plus Local Orbitals Program for Calculating Crystal Properties,* Karlheinz Schwarz/Techn. Universität Wien, Wien, Austria, (2001)
[37] A.F.Kemper, M.M.Korshunov, T.P.Devereaux, J.N.Fry, H-P.Cheng, P.J.Hirschfeld, Phys. Rev B 83, 184516 (2011)
[38] M.Matusiak, Z.Bukowski, J.Karpinski, Phys. Rev. B 83, 224505 (2011)
[39] M.Tropeano, I.Pallecchi, M.R.Cimberle, C.Ferdeghini, G.Lamura, M.Vignolo, A.Martinelli, A.Palenzona, M.Putti, Supercond. Sci. Technol. 23, 054001 (2010)
[40] J.Nilsson, A.H.Castro Neto, F.Guinea, N.M.R.Peres, Phys. Rev. B 78, 045405 (2008)
[41] F.Schedin, A.K.Geim, S.V.Morozov, E.W.Hill, P.Blake, M.I.Katsnelson, K.S.Novoselov, Nature Mat. 6, 652 (2007)